\documentclass[twocolumn,showpacs,preprintnumbers,amsmath,amssymb,superscript address, revsymb,prb]{revtex4}

\usepackage{graphicx}
\usepackage{mathptmx}
\usepackage{dcolumn}
\usepackage{bm}

\newcolumntype{.}{D{.}{\cdot}{3.10}}

\newcommand{\greekvect}[1]{{\bm {#1}}}
\newcommand{\vect}[1]{{\bf {#1}}}

\begin{document}

\title{Ultracold atoms at unitarity within quantum Monte Carlo}


\author{Andrew J. Morris\footnote{Email: ajm255@cam.ac.uk.}}
\affiliation{Theory of Condensed Matter Group, Cavendish Laboratory,
University of Cambridge, J. J. Thomson Avenue, Cambridge CB3 0HE,
United Kingdom.}

\author{P. L\'opez R\'ios} \affiliation{Theory of Condensed Matter
Group, Cavendish Laboratory, University of Cambridge, J. J. Thomson
Avenue, Cambridge CB3 0HE, United Kingdom.}

\author{R. J. Needs} \affiliation{Theory of Condensed Matter Group,
Cavendish Laboratory, University of Cambridge, J. J. Thomson Avenue,
Cambridge CB3 0HE, United Kingdom.}

\date{\today{}}

\begin{abstract} 
Variational and diffusion quantum Monte Carlo (VMC and DMC)
calculations of the properties of the zero-temperature fermionic gas
at unitarity are reported.  The ratio of the energy of the interacting
to the non-interacting gas for a system of 128 particles is calculated
to be 0.4517(3) in VMC and 0.4339(1) in the more accurate DMC method.
The spherically-averaged pair-correlation functions, momentum
densities, and one-body density matrices are very similar in VMC and
DMC, but the two-body density matrices and condensate fractions show
some differences.  Our best estimate of the condensate fraction of
0.51 is a little smaller than values from other quantum Monte Carlo
calculations.
\end{abstract}



\maketitle

\section{Introduction} 
\label{sec:Introduction}
Superfluid pairing in ultracold trapped atoms has been the subject of
much experimental and theoretical
work.\cite{Ketterle:RevModPhys:2002,Cornell:RevModPhys:2002,Kohler:RevModPhys:2006,Giorgini:RevModPhys:2008,Bloch:RevModPhys:2008}
The range of the inter-atomic interaction in a dilute atomic Fermi gas
is much smaller than the average distance between the atoms, and only
the $s$-wave scattering length $a$ is relevant.  The only relevant
dimensionless coupling parameter is then $1/(ak_{\rm F})$, where
$k_{\rm F}$ is the Fermi wave vector of the gas.  The scattering
length can be altered by applying a magnetic field and, by using
Fano-Feshbach
resonances,\cite{Fano:PR:1961,Feshbach:AnnPhys:1962,Inouye:Nature:1998,Courteille:PRL:1998}
$1/(ak_{\rm F})$ may be varied from large negative to large positive
values.  When the interaction is weakly attractive and $k_{\rm F}$ is
sufficiently small, $1/(ak_{\rm F}) \rightarrow -\infty$, and the gas
is in the Bardeen-Cooper-Schrieffer (BCS) superfluid regime.  When
$1/(ak_{\rm F}) \rightarrow +\infty$, the molecules are tightly bound
and the system forms a Bose-Einstein condensate (BEC)\@.  The behavior
in the intermediate regime changes smoothly with $1/(ak_{\rm F})$ and
at unitarity, where $|1/(ak_{\rm F})| \rightarrow 0$, a smooth
crossover between BCS-like and BEC-like behavior occurs.

At unitarity the scattering length becomes larger than the
inter-particle distance and the only energy scale is $k_{\rm F}^2$ (we
consider a particle mass of unity).\cite{units} The ground state
energy $E_0$ can therefore be written as
\begin{eqnarray}
E_0 = \xi \frac{3}{10} {k_{\rm F}^2} ,
\end{eqnarray}
where the factor of 3/10 is chosen so that $\xi$ is the fraction of
the energy of the non-interacting Fermi gas at the same density.  A
number of experimental and theoretical determinations of the universal
parameter $\xi$ have been reported.  In each case the parameter was
found to be smaller than unity, showing that the interactions are
attractive at unitarity.

In this paper we report calculations of the energy, pair correlation
functions (PCFs), momentum density and the one- and two-body density
matrices of the Fermi gas at unitarity.  We use the zero-temperature
variational and diffusion quantum Monte Carlo
methods\cite{Foulkes:RMP:2001} (VMC and DMC), as have been used in
previous
studies.\cite{Carlson:PRL:2003,Astrakharchik:PRL:2004,Chang:PRA:2004,Astrakharchik:PRL:2005,Chang:PRL:2005,Carlson:PRL:2005,Gezerlis:PRL:2009}
Our study differs from earlier ones mainly in the construction of the
trial wave functions, the larger system size used, and in studying the
dependence of the energy on the particle density.  Other quantum Monte
Carlo (QMC) methods have been used to study ultracold atomic systems
at finite temperatures.\cite{Akkineni:PRB:2007,Burovski:PRL:2008}

The rest of this paper is set out as follows.  In Sec.\
\ref{sec:The_Hamiltonian} we discuss the Hamiltonian used to model the
atomic Fermi gas. An introduction to the VMC and DMC methods is given
in Sec.\ \ref{subsec:VMC_and_DMC_methods}, and specific points pertaining to
Fermi atomic gases are described in Sec.\
\ref{subsec:VMC_and_DMC_calculations}.  Our trial wave functions are
described in Sec.\ \ref{sec:Trial_Wave_Functions} and important
parameters of the DMC calculations are discussed in Sec.\
\ref{sec:DMC_calculations}. Our results are reported and discussed in
Sec.\ \ref{sec:Results} and the conclusions of our study are
summarized in Sec.\ \ref{sec:Conclusions}.


\section{The Hamiltonian}
\label{sec:The_Hamiltonian}

The Hamiltonian takes the form
\begin{equation}
\hat{H} = - \frac{1}{2} \sum_i \nabla^2_i + \sum_{i<j} v(r_{ij}) ,
\end{equation}
where $v(r_{ij})$ is the interaction potential.  We use face-centered
cubic (fcc) simulation cells subject to periodic boundary conditions.
We wish to study the system with a delta-function potential, but this
is difficult to sample using Monte Carlo methods. We have instead used
the P\"{o}schl-Teller interaction which, on resonance, is given by
\begin{equation}
v({r}_{ij}) = -\frac{2\mu}{\mathrm{cosh}^2\left(\mu {r}_{ij}\right)},
\end{equation}
where ${r}_{ij}$ is the distance between particles $i$ and $j$, and
$2/\mu$ is the effective width of the potential well.  Since the
inter-particle interaction is very short-ranged, particles of the same
spin are kept apart by the antisymmetry of the wave function, and the
interaction between them is negligible for the well widths used here
and would be precisely zero for the delta-function potential.  We
therefore set the interaction between particles of the same spin to
zero, as has been done in previous calculations.  The
P\"{o}schl-Teller interaction has been used in previous QMC
calculations,\cite{Carlson:PRL:2003,Carlson:PRL:2005} and we prefer it
to the square-well which has also been used in QMC
calculations,\cite{Astrakharchik:PRL:2004,Astrakharchik:PRL:2005}
because its smoothness aids Monte Carlo sampling.  In all of our QMC
calculations reported here we have used $\mu=12$.

The particle density is $k_{\rm F}^3/(3\pi^2)$, but we report
densities in terms of the $r_s$ parameter, which is the radius of a
sphere containing one atom on average, and $r_s =
(9\pi/4)^{1/3}/k_{\rm F}$.  For most of our calculations we have used
a density parameter $r_s=1$, so that $\mu r_s \gg 1$, as required for
dilute conditions, although in Sec.\
\ref{subsec:Varying_the_particle_density} we report some
investigations of the effect of increasing $r_s$.

\section{QMC methods}
\label{sec:QMC_methods}

\subsection{VMC and DMC methods}
\label{subsec:VMC_and_DMC_methods}

In VMC the energy is calculated as the expectation value of the
Hamiltonian with an approximate many-body trial wave function.  In the
more accurate DMC method the ground state energy is obtained by
evolving the wave function in imaginary time so that it decays towards
the ground state.  Projector methods such as DMC suffer from a fermion
sign problem, which is evaded by making the fixed-node approximation,
and importance sampling is introduced to reduce the statistical noise.
The importance-sampled fixed-node fermion DMC algorithm was first used
by Ceperley and Alder to study the electron
gas.\cite{Ceperley:PRL:1980}

The key quantity in VMC and DMC calculations is the trial wave
function, which controls both the accuracy that can be obtained and
the statistical efficiency of the computation.  In VMC the accuracy of
the energy estimate depends on the whole of the trial wave function,
while in DMC it depends only on the form of its nodal surface, as the
DMC algorithm (in principle) gives the lowest energy compatible with
the fixed nodal surface.  In practice the DMC energy estimate also
shows some dependence on the timestep used and a very weak dependence
on the population of particle configurations.  Improving the trial
wave function tends to reduce these biases and improve the statistical
efficiency of the calculations.

The VMC algorithm generates particle configurations distributed
according to 
\begin{eqnarray}
\label{eq:VMC_distribution}
p_{\rm V}({\bf R}) = \frac{\Psi_{\rm T}^2({\bf R})}{\int \Psi_{\rm
T}^2({\bf R}^{\prime}) \, d{\bf R}^{\prime}} \;,
\end{eqnarray}
where $\Psi_{\rm T}$ is a real trial wave function and $\vect{R}$ is
the 3$N$-dimensional vector of the coordinates of the $N$ particles.
DMC generates configurations distributed according to
\begin{eqnarray}
\label{eq:DMC_distribution}
p_{\rm D}({\bf R}) = \frac{\Psi_{\rm T}({\bf R})\Phi({\bf R})}{\int
\Psi_{\rm T}({\bf R}^{\prime}) \Phi({\bf R}^{\prime}) \, d{\bf R}^{\prime}}
\;,
\end{eqnarray}
where $\Phi({\bf R})$ is the DMC wave function.  The total energy in
both VMC and DMC is calculated from
\begin{eqnarray}
\label{eq:local_energy}
E = \int p({\bf R}) E_{\rm L}({\bf R}) \, d{\bf R} \;,
\end{eqnarray}
where $E_{\rm L}({\bf R}) = \Psi_{\rm T}^{-1} \hat{H} \Psi_{\rm T}$ is
the local energy, and $p = p_{\rm V}$ in VMC and $p = p_{\rm D}$ in
DMC.

DMC expectation values of operators which do not commute with the
Hamiltonian depend on the entire trial wave function, not just its
nodal surface.  To reduce this bias, at the expense of increasing the
noise, one can use the extrapolation
approximation,\cite{Ceperley:book:1986}
\begin{eqnarray}
\label{eq:extrapolation}
\left<\hat{A}\right> \simeq 2\left<\hat{A}\right>_{\rm
DMC}-\left<\hat{A}\right>_{\rm VMC} + {\cal O} \left[(\Psi_{\rm T} -
\Phi)^2\right],
\end{eqnarray}
where $\left<\hat{A}\right>_{\rm DMC}$ and $\left<\hat{A}\right>_{\rm
VMC}$ are the DMC and VMC expectation values of operator $\hat{A}$,
respectively.  The quantity $\left<\hat{A}\right>_{\rm
DMC}-\left<\hat{A}\right>_{\rm VMC}$ gives a measure of the accuracy
of the trial wave function.

\subsection{VMC and DMC calculations for Fermi atomic gases}
\label{subsec:VMC_and_DMC_calculations}

The construction of accurate trial wave functions for Fermi atomic
gases is not straightforward.  The variation of the wave function must
be described very accurately at small inter-particle separations where
the interaction potential varies very rapidly.  The binding energy of
an isolated molecule is vanishingly small at resonance, but the exact
value of $\xi$ for the gas is certainly smaller than the BCS
mean-field value of 0.59, and therefore the interactions between
molecules are very important at unitarity.
The exact wave function for an isolated pair of opposite spin
fermionic atoms at resonance decays as the inverse distance between
the particles, and we must describe the deviations from this behavior
in the gas phase.  We conclude that it is necessary to provide a good
description of both the long- and short-ranged behavior of the pairing
function to obtain accurate results for the system at unitarity.

The simulations are performed with a finite number of particles, and
we wish to obtain results which accurately reflect those that would be
obtained with an infinite number.  A great deal of experience has been
gained in performing QMC calculations for electron gases, and the
finite size effects are greatly reduced by averaging the energies
obtained at different wave vectors\cite{Lin:PRE:2001} (``twist
averaging'') and extrapolating the averaged energies to the infinite
system limit.  The QMC studies of ultracold atoms reported so far have
not employed very large numbers of particles and have not used twist
averaging, and it is not clear whether the results are converged with
respect to system size.  Although we have not used twist averaging in
the calculations reported here, we have used systems with 128
particles, which is approximately twice the largest number used in
previous DMC
calculations.\cite{Carlson:PRL:2003,Astrakharchik:PRL:2004,Chang:PRA:2004,Astrakharchik:PRL:2005,Chang:PRL:2005,Carlson:PRL:2005,Gezerlis:PRL:2009}

Another problem is that we really want the solution for a
delta-function potential, but for computational reasons we use a well
of finite width.  The ground state of the many-particle system with a
delta-function potential is a molecular gas because bound states with
more than two particles cannot exist in this case but, with a finite
well-width, clusters of particles can form at high densities.  In
practice this instability of the gas phase occurs only for densities
where $\mu r_s \ll 1$, and we work at much lower densities.
Nevertheless, it is clear that the results of QMC calculations will
depend on both the well width and the particle density.

\subsection{Trial wave functions}
\label{sec:Trial_Wave_Functions}

We used a singlet-pairing BCS-like wave function consisting of a
determinant of identical pairing orbitals, $\varphi \left( r_{ij}
\right)$, each of which is a function of the separation of an up- and
a down-spin particle.  The pairing orbital is represented by a sum of
polynomial terms,
\begin{eqnarray}
\label{eq:pairing_orbital}
\varphi \left( r_{ij} \right) & = & \left ( \frac{L_{\mathrm
P}-{r}_{ij}}{L_{\mathrm P}} \right )^{3} \sum_{n=0}^{N_{\mathrm{P}}}
\gamma_n {r}_{ij}^n .
\end{eqnarray}
We set $\varphi$ to zero for $r_{ij}$ greater than the cutoff length
$L_{\rm P}$.  The third power in Eq.\ (\ref{eq:pairing_orbital}) was
chosen to ensure that the local energy is continuous at $L_{\rm P}$.
The $\gamma_n$ are optimizable parameters, but the value of $\gamma_1$
is determined by the condition that $\varphi$ is cuspless at the
origin.  We tested various values of $N_{\rm P}$ and chose $N_{\rm
P}=4$ for the results presented here.
The optimized value of the cutoff length of $L_{\rm P} = 3.5$ a.u.\ is
3.5 times the average distance between particles.  We also tested
orbitals represented by linear combinations of Gaussian orbitals,
linear combinations of plane waves, and linear combinations of
Gaussian orbitals, plane waves, and polynomials.  Gaussian orbitals
also appeared to be a useful choice, but we chose the polynomial basis
as the orbitals and their derivatives can be evaluated much more
rapidly.

The determinant of pairing orbitals is multiplied by a Jastrow factor
of the form $e^J$, with
\begin{equation}
J(\vect{r}_{ij}) = \sum_{s=1}^{N_{\mathrm{s}}} \lambda_s
\sum_{\vect{G} \in {\mathrm s}}\exp \left \{ \mathrm{i}\vect{G} \cdot
\vect{r}_{ij} \right \} + \left ( \frac{L_{\mathrm
J}-{r}_{ij}}{L_{\mathrm J}} \right )^{3} \sum_{n=0}^{N_{\mathrm{J}}}
\theta_n {r}_{ij}^n ,
\end{equation}
where $N_{\rm s}$ is the number of stars of symmetry-related
reciprocal lattice vectors $\vect{G}$, and $N_{\rm J}$ is the order of
the polynomial. The $\lambda_s$ and $\theta_n$ are optimizable
parameters, although $\theta_1$ is determined by the condition that
$J$ is cuspless at the origin, and we set the polynomial part of $J$
to zero for $r_{ij} > L_{\rm J}$.\cite{Drummond:PRB:2004} After some
testing we chose $N_{\rm s}=4$ an $N_{\rm J}=8$, and we used an
optimized value of $L_{\rm J}=0.86$ a.u.

We also applied backflow a transformation to the determinant of
pairing orbitals.\cite{Feynman:PR:1954,Feynman:PR:1956} The particle
coordinates $\vect{r}_i$ are replaced by collective coordinates
$\vect{x}_i(\vect{R})=\vect{r}_i+\greekvect{\zeta}_i(\vect{R})$, where
$\greekvect{\zeta}_i(\vect{R})$ is the backflow displacement of
particle $i$, which depends on all the particle positions.  The
backflow displacement is given by
\begin{equation}
\greekvect{\zeta}_{i}(\vect{R}) = \sum_{j \ne i} {\eta}_{ij}({r}_{ij})
\vect{r}_{ij}.
\end{equation}
We have used the form
\begin{equation}
{\eta}_{ij}({r}_{ij}) = \left ( \frac{L_{\mathrm B}-r_{ij}}
{L_{\mathrm B}} \right )^3 \sum_{n=0}^{N_{\mathrm{B}}} \rho_n
{r}_{ij}^n,
\end{equation}
where $L_{\rm B}$ is a cutoff length and $N_{\rm B}$ is the order of
the polynomial in the backflow expansion and the $\rho_n$ are
optimizable parameters, with $\rho_1$ determined by the condition that
$\eta$ is cuspless at the origin.\cite{Lopez-Rios:PRE:2006} We chose
$N_{\rm B}=5$ and used an optimized value of $L_{\rm B}=1.04$ a.u.

The wave functions were optimized within a VMC procedure by minimizing
the mean absolute deviation of the local energies from their median
value.  We found this approach to be superior to variance minimization
schemes.\cite{Kent:PRB:1999,Drummond:PRB:2005} The polynomial term in
the Jastrow factor is much more important than the plane-wave part.
The Jastrow and backflow functions can, in principle, operate between
both parallel and anti-parallel spin particles, although the
correlation effects between the parallel-spin particles beyond the
exchange interaction already included in the determinant are small.
We did, however, find a small lowering of the VMC energy when we
allowed the plane wave parameters in the Jastrow factor to be non-zero
for parallel-spin particles.  We did not include parallel-spin terms
for the polynomials.  The wave function contains a total of 28
variable parameters.

\subsection{QMC calculations}
\label{sec:DMC_calculations}

We used the \textsc{CASINO} code\cite{casino} for all of our QMC
calculations.  We performed some test calculations with 32 and 64
particles, but all of our results reported in this paper were obtained
with 128 particles.
We used a time step of 0.015 a.u.\ for all the DMC results presented
in this paper.  Test calculations using a timestep of 0.03 a.u.\ did
not change the total energy within the statistical error bars
achieved.  We used a mean population of 3200 configurations, and tests
indicated that the population control bias with this number of
configurations is negligible.

\section{Results}
\label{sec:Results}

\subsection{Total energy and the $\xi$ parameter}
\label{subsec:Total_Energy_and_the_xi_parameter}

\begin{figure}
\includegraphics*[width=8cm]{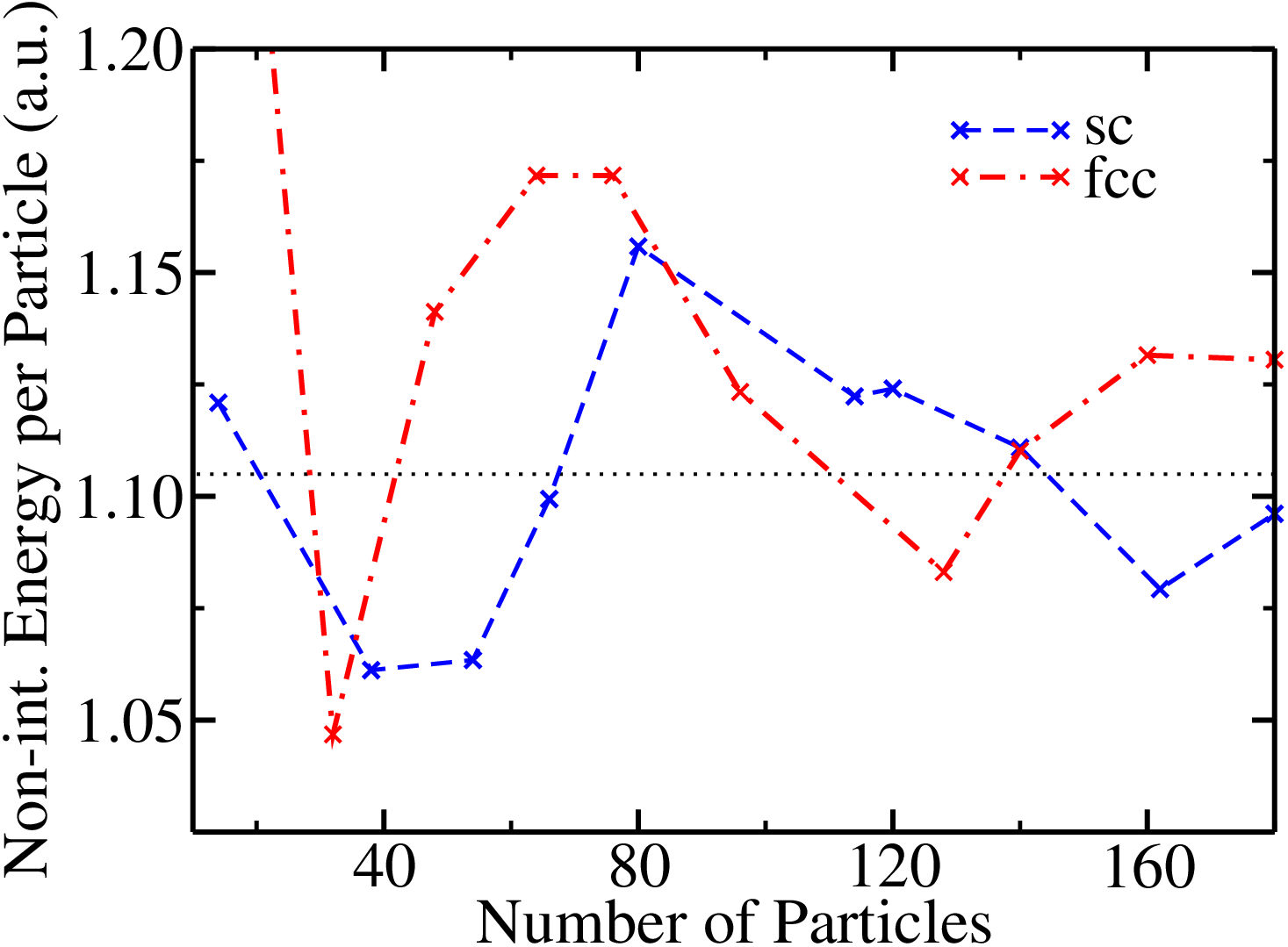}
\caption[]{(Color online) Energy per particle of the non-interacting
system at $r_s=1$ for different particle numbers, for the sc and fcc
lattices.  The dotted line at 1.10495 a.u.\ shows the exact energy of
the infinite system.}
\label{Fig:non-int}
\end{figure}

When evaluating the $\xi$ parameter it is not immediately obvious
whether to use the non-interacting energy $E_{\rm NI}$ of the finite
system studied, the infinite system, or some other value.  Energies of
non-interacting systems for various particle numbers and fcc and
simple cubic (sc) simulation cells are shown in Fig.\
\ref{Fig:non-int}.  $E_{\rm NI}$ oscillates in an irregular manner
about the infinite system value as the particle number $N$ is
increased and converges rather slowly with $N$.  Earlier DMC
calculations of $\xi$ used
sc cells and particle numbers from $N=14$ to
66.\cite{Carlson:PRL:2003,Astrakharchik:PRL:2004,Chang:PRL:2005,Carlson:PRL:2005}
The error in $E_{\rm NI}$ for the 66-particle system is quite small
(0.5\%), although it increases to nearly 5\% for the sc cell with 80
particles.  The rate of convergence for
the fcc cell is similar to that for the sc cell.

The convergence of the interacting energy with system size has not
been well-studied, but using more particles 
is expected to improve the results.  The momentum distribution of the
non-interacting system has a discontinuity at the Fermi momentum,
while that of the interacting system is smooth, see Sec.\
\ref{subsec:Momentum_Density_and_Density_Matrices}, so that the
kinetic energy of the non-interacting system varies rapidly in
k-space, leading to rapid fluctuations in the kinetic energy with
system size.
It therefore seems likely that the interacting energy will converge
faster with system size than the non-interacting energy.

Our values of $\xi$, and those from other calculations and some
experiments, are given in Table \ref{table:chi}.  The DMC energy is
bounded from above by the VMC energy and from below by the exact
energy and, as expected, the VMC values of $\xi$ are a little larger
than the DMC ones.  The results reported in Table \ref{table:chi} were
obtained using the value of $E_{\rm NI}$ for the finite system studied
of 1.08307 a.u.\ per particle, compared with the infinite-system
result of 1.10495 a.u.\ per particle.  As discussed above, one might
argue that it is more appropriate to calculate $\xi$ using the
infinite-system value of $E_{\rm NI}$, as the interacting energy is
expected to converge faster with system size than the non-interacting
energy.  Using the infinite-system non-interacting energy gives values
of $\xi=0.4428(1)$ in VMC, and $\xi=0.4253(1)$ in DMC, which are about
2\% lower than those reported in Table \ref{table:chi}.
Our DMC value of $\xi$ is very similar to previous
ones.\cite{Carlson:PRL:2003,Astrakharchik:PRL:2004,Chang:PRL:2005,Carlson:PRL:2005}

\begin{table}
\begin{tabular}{l|c}
\hline\hline
Method                           & $\xi$       \\
\hline
Exp.\cite{Bartenstein:PRL:2004}  & 0.32(10)    \\
Exp.\cite{Kinast:Science:2005}   & 0.51(4)     \\
Exp.\cite{Partridge:Science:2006}& 0.46(5)     \\
Theor.\cite{Perali:PRL:2004}     & 0.455       \\
Theor.\cite{Nishida:PRA:2009}    & 0.360(20)   \\
DMC\cite{Carlson:PRL:2003}       & 0.44(1)     \\
DMC\cite{Astrakharchik:PRL:2004} & 0.42(1)     \\
DMC\cite{Chang:PRL:2005}         & 0.414(5)    \\
DMC\cite{Carlson:PRL:2005}       & 0.42(1)     \\
\hline
VMC This work                    & 0.4517(3)   \\
DMC This work                    & 0.4339(1)   \\
\hline\hline
\end{tabular}
\caption[Values of $\xi$]{Values of the universal parameter $\xi$ from
experiments and theory.  The error bars given in brackets for the
current work are purely statistical errors, and they do not account
for biases from finite size effects or fixed-node errors or other
sources.}
\label{table:chi}
\end{table}

\subsection{Pair correlation functions}
\label{subsec:Pair_Correlation_Functions}

\begin{figure}
\includegraphics*[width=8cm]{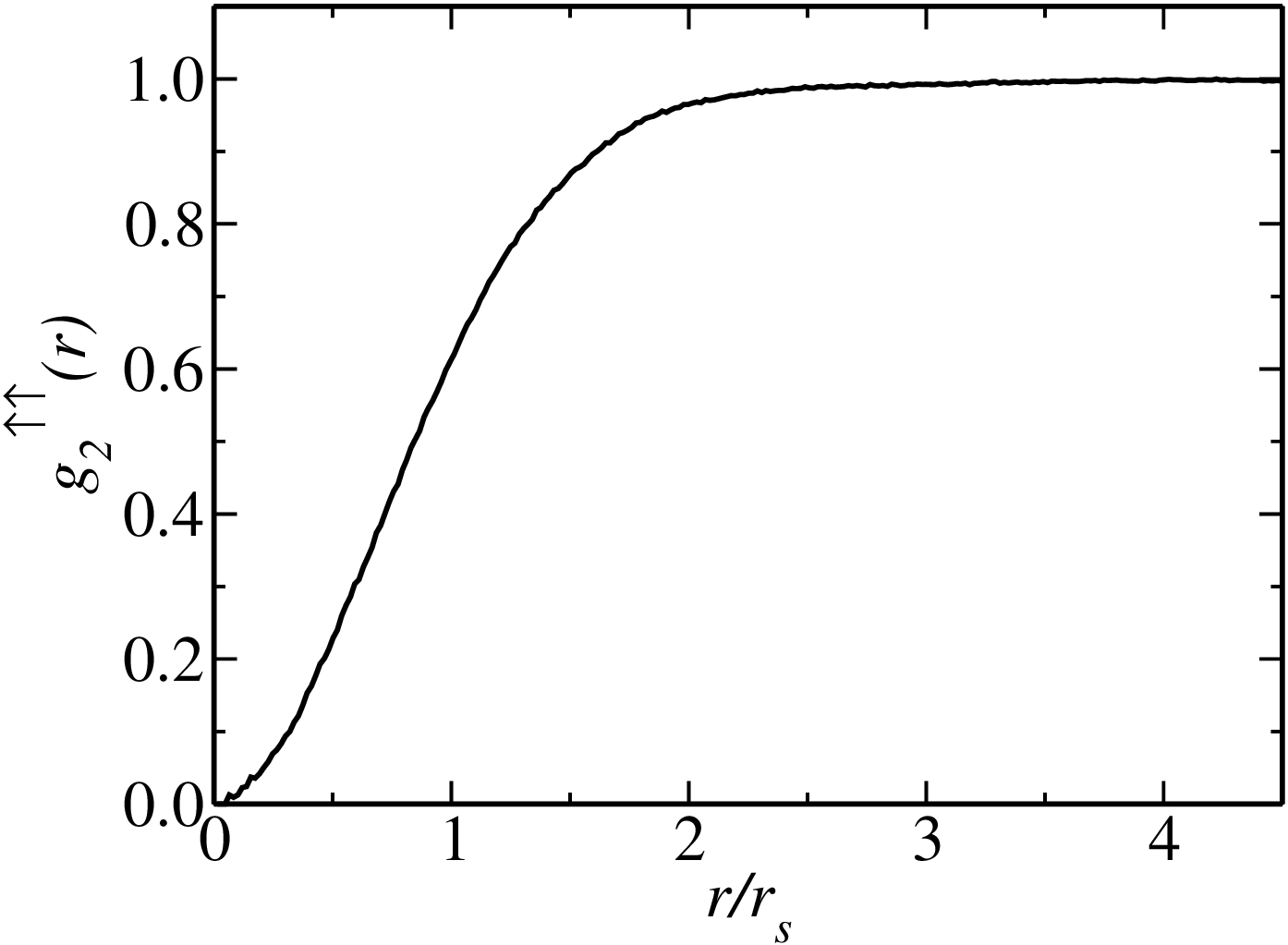}
\caption[]{Raw data for the parallel-spin PCF.}
\label{Fig:parallel_pcf}
\end{figure}

\begin{figure}
\includegraphics*[width=8cm]{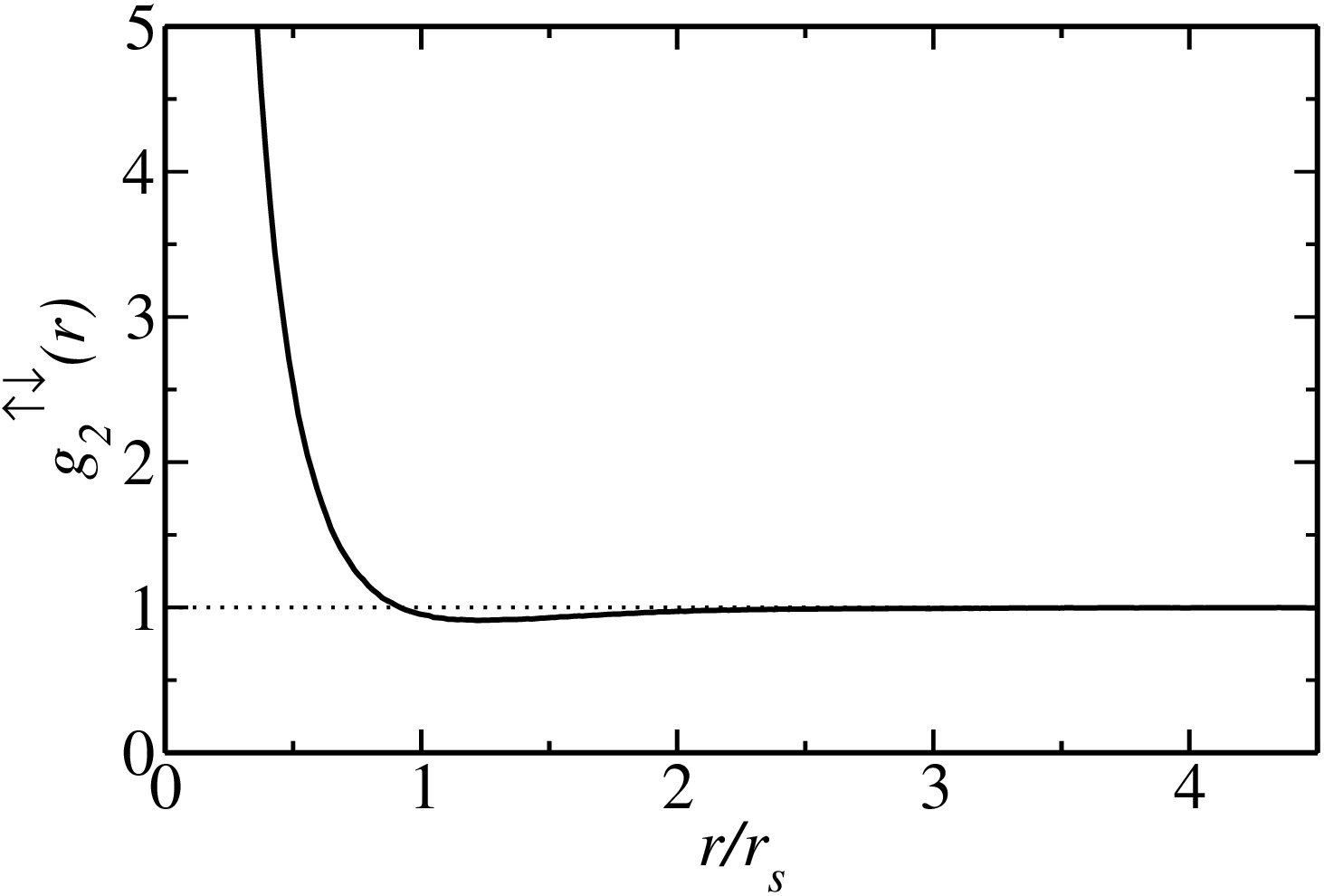}
\caption[]{Raw data for the anti-parallel-spin PCF.}
\label{Fig:anti-parallel_pcf}
\end{figure}

We evaluated the spatially and rotationally averaged pair correlation
functions (PCFs) for the parallel and anti-parallel spin pairs, which
are shown in Figs.\ \ref{Fig:parallel_pcf} and
\ref{Fig:anti-parallel_pcf}.  
The difference in our VMC and DMC results was negligible on the scales
of Figs.\ \ref{Fig:parallel_pcf} and \ref{Fig:anti-parallel_pcf}, and
consequently the effect of using the extrapolation of Eq.\
(\ref{eq:extrapolation}) is negligible.  
Figures \ref{Fig:parallel_pcf} and \ref{Fig:anti-parallel_pcf} show
the DMC data itself, not fits to the data.  The noise in the data is
very small, but can just be resolved in Fig.\ \ref{Fig:parallel_pcf}
at small $r/r_s$, where the number of counts is small.  The
parallel-spin PCF shown in Fig.\ \ref{Fig:parallel_pcf} shows a hole
largely confined to the region $r/r_s < 2$, which is essentially an
exchange hole, and the PCF is very similar to the non-interacting
(Hartree-Fock) result.  This is consistent with the fact that we found
only a very weak parallel-spin Jastrow factor.  The anti-parallel-spin
PCF (Fig.\ \ref{Fig:anti-parallel_pcf}) shows a very strong
enhancement for $r/r_s < 1$ arising from the pairing.  The behavior at
small $r/r_s$ is not shown as it depends strongly on the well width.
The anti-parallel PCF dips below unity in the region $1 < r/r_s < 2$.
The PCFs are similar to those reported in Fig.\ 3 of Astrakharchik
\textit{et al.}\cite{Astrakharchik:PRL:2004} and Fig.\ 1 of Chang and
Pandharipande.\cite{Chang:PRL:2005}


\subsection{Momentum density and density matrices}
\label{subsec:Momentum_Density_and_Density_Matrices} 

The one-body density-matrix (OBDM) may be written as
\begin{equation}\label{eq:1bdm_def}
\rho_\alpha^{(1)}({\bf r}_1;{\bf r}_1^\prime) = N_\alpha \frac {\int
p({\bf R}) \frac {\Psi({\bf r}_1^\prime) } {\Psi({\bf r}_1) } \, d{\bf
r}_2 \dots d{\bf r}_N } {\int p({\bf R}) \, d{\bf R} } \;,
\end{equation}
and the two-body density-matrix (TBDM) as
\begin{equation}\label{eq:2bdm_def}
\rho_{\alpha\beta}^{(2)}({\bf r}_1,{\bf r}_2;{\bf r}_1^\prime, {\bf
r}_2^\prime) = N_\alpha (N_\beta-\delta_{\alpha\beta}) \frac {\int
p({\bf R}) \frac {\Psi({\bf r}_1^\prime,{\bf r}_2^\prime) } {\Psi({\bf
r}_1,{\bf r}_2) } \, d{\bf r}_3 \dots d{\bf r}_N } {\int p({\bf R}) \,
d{\bf R} } \;,
\end{equation}
where $p({\bf R})$ is the VMC or DMC probability distribution of Eqs.\
(\ref{eq:VMC_distribution}) or (\ref{eq:DMC_distribution}), ${\bf
r}_1$ and ${\bf r}_1^{\prime}$ are $\alpha$-spin particle coordinates,
${\bf r}_2$ and ${\bf r}_2^{\prime}$ are $\beta$-spin particle
coordinates, and $N_{\alpha} = N_{\beta} = N/2$ is the number of
particles of each spin type.

\begin{figure}
\includegraphics*[width=8cm]{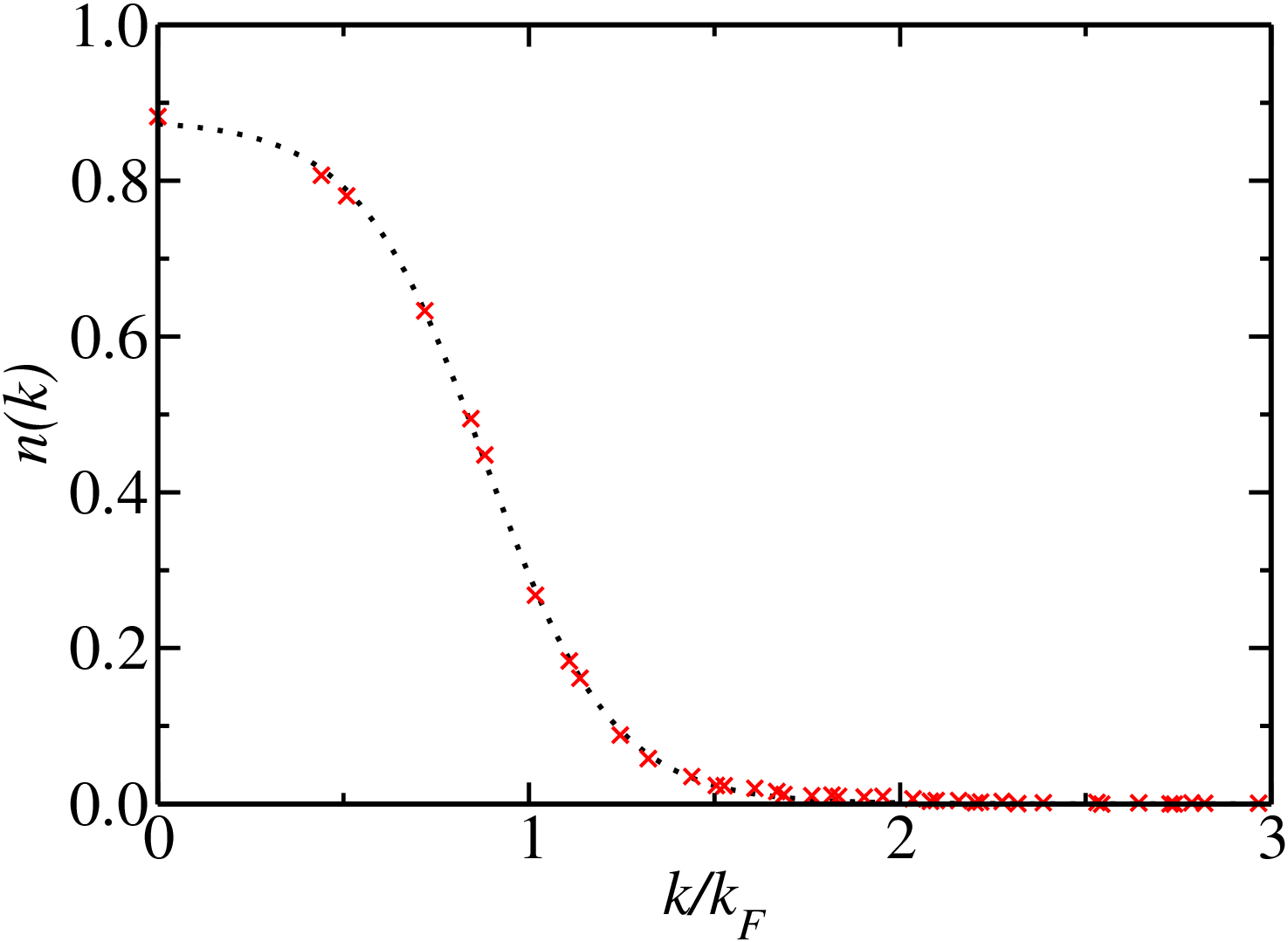}
\caption[]{(Color online)  Momentum density.  The raw DMC data are shown
(crosses) and the dotted line is a fit to this data.}
\label{Fig:Gamma_DMC}
\end{figure}

We have evaluated the translationally and rotationally averaged
density matrices, which we denote by $\rho_\alpha^{(1)}(r)$ and
$\rho_\alpha^{(2)}(r)$, respectively.  The momentum density $n(k)$ is
the Fourier transform of $\rho_\alpha^{(1)}(r)$, but we evaluate it
directly in Fourier space, which is a somewhat better numerical
approach.  Our data for $n(k)$ at unitarity are broadly similar to
those presented in Fig.\ 2 of Ref.\ \onlinecite{Carlson:PRL:2003} and
Fig.\ 2 of Ref.\ \onlinecite{Astrakharchik:PRL:2005}, although in
detail there are some differences.  Our data show a monotonic decrease
of $n(k)$ with increasing momentum, in common with the results of
Ref.\ \onlinecite{Astrakharchik:PRL:2005}, but in conflict with Ref.\
\onlinecite{Carlson:PRL:2003}, which show a peak below the Fermi
momentum.  Carlson \textit{et al.}\cite{Carlson:PRL:2003} show VMC and
DMC data for 14 and 38 particles.  The VMC and DMC data are in good
agreement above $k_{\rm F}$, but below $k_{\rm F}$ there are
substantial differences. The differences between our VMC and DMC data
are very small, and would be barely visible on the scale of Fig.\
\ref{Fig:Gamma_DMC}. This suggests that our trial wave functions are
superior to those of Ref.\ \onlinecite{Carlson:PRL:2003}.
Astrakharchik \textit{et al.}\cite{Astrakharchik:PRL:2005} only report
DMC data in the form of a fit, rather than giving the calculated
values.  Our momentum density is a little closer to the BCS form than
the curve shown in Ref.\ \onlinecite{Astrakharchik:PRL:2005}.

\begin{figure}
\includegraphics*[width=8cm]{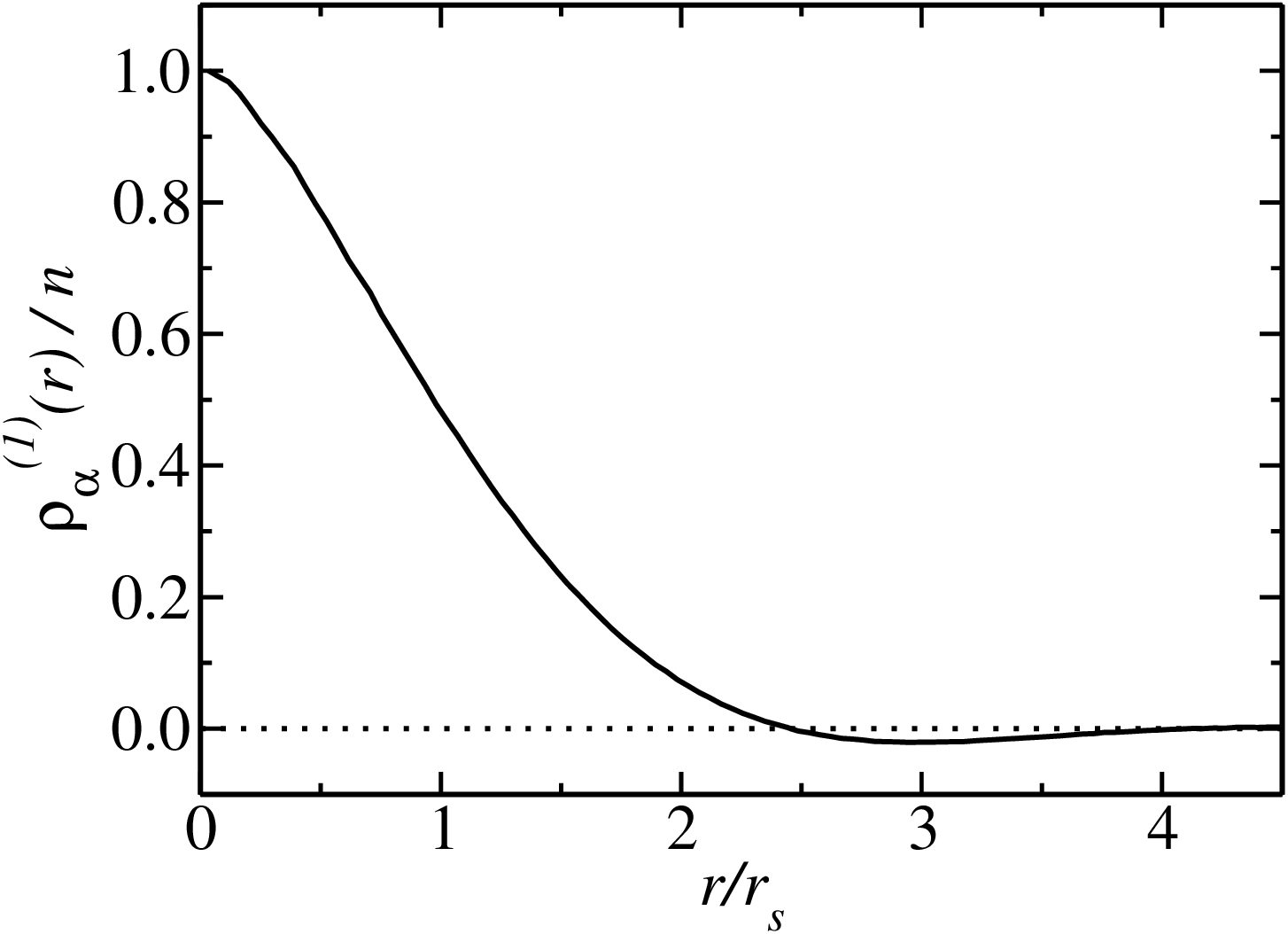}
\caption[]{Raw DMC data for the one-body density-matrix.}
\label{Fig:obdm}
\end{figure}

The OBDM is shown in Fig.\ \ref{Fig:obdm}. Again, the VMC and DMC data
are virtually indistinguishable so that extrapolation is unnecessary.
Our calculated OBDM is very similar to that shown in Fig.\ 1 of Ref.\
\onlinecite{Astrakharchik:PRL:2005}.

\begin{figure}
\includegraphics*[width=8cm]{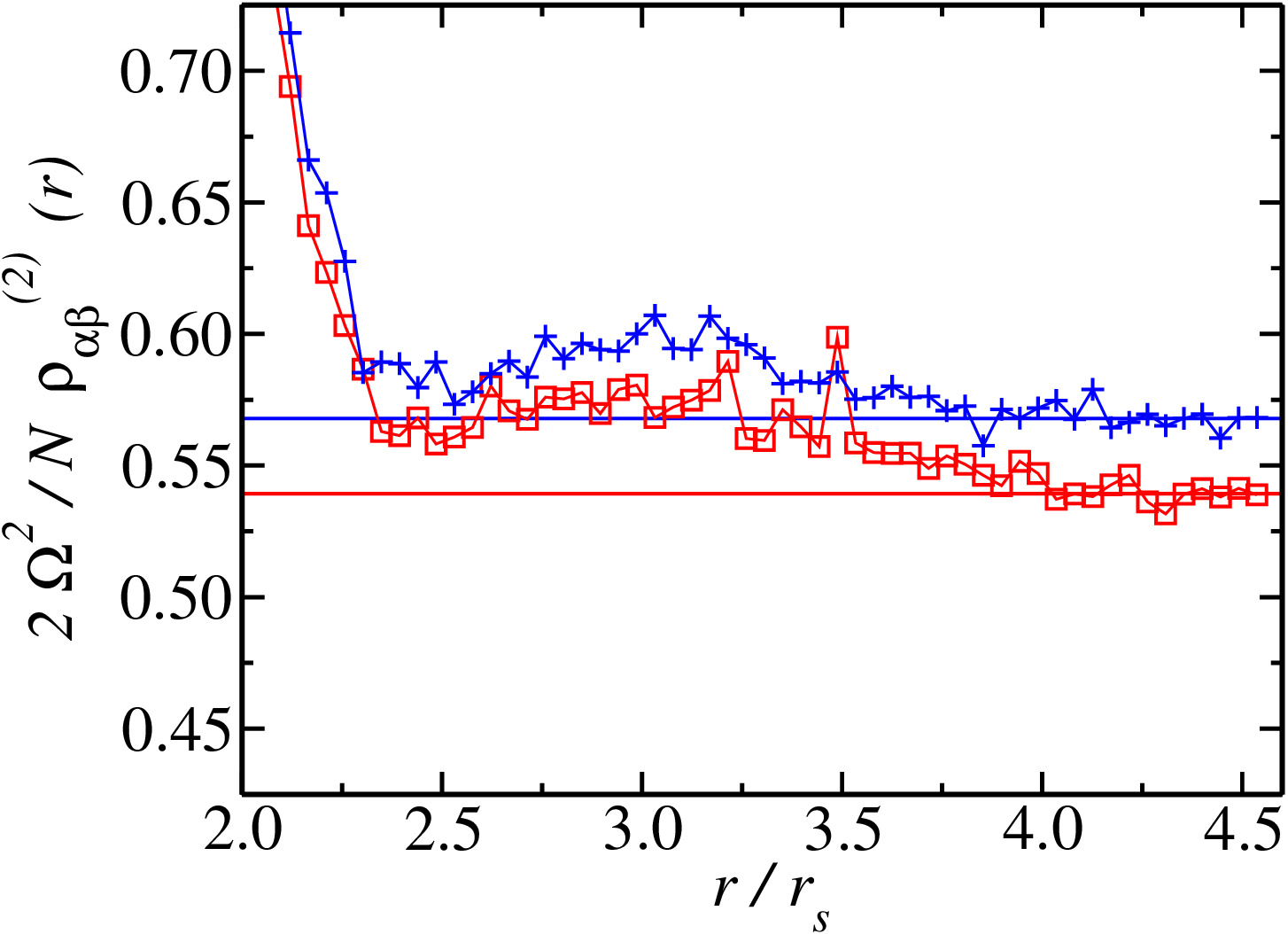}
\caption[]{(Color online) The two-body density-matrix.  The VMC data
are shown as plus signs and the DMC data as squares.  The horizontal
lines shows estimates of the asymptotic behavior which give the
condensate fraction.}
\label{Fig:tbdm}
\end{figure}

The condensate fraction $c$ is related to the translationally and
rotationally averaged TBDM by
\begin{equation}
c = \frac{2\Omega^2}{N} \lim_{r\rightarrow \infty}
\rho_{\alpha\beta}^{(2)}(r),
\end{equation}
where $\Omega$ is the volume of the simulation cell and $N/2$ is the
number of pairs of particles in the system.  VMC and DMC data for the
TBDM are shown in Fig.\ \ref{Fig:tbdm}.  In this case we find some
differences between the VMC and DMC results.
We obtained a condensate fraction of $c=0.57$ in VMC and $c=0.54$ in
DMC, see Fig.\ \ref{Fig:tbdm}, so that the value obtained using the
extrapolation of Eq.\ (\ref{eq:extrapolation}) is 0.51. Our DMC values
are lower than those of $0.61(2)$ for 38 particles and $0.57(2)$ for
66 particles reported by Astrakharchik \textit{et
al.}\cite{Astrakharchik:PRL:2005}

\subsection{Varying the particle density}
\label{subsec:Varying_the_particle_density}

As mentioned in Sec.\ \ref{sec:The_Hamiltonian}, we require $\mu r_s
\gg 1$ for dilute conditions.  This can be satisfied by, for example,
fixing $r_s$ and choosing the effective width of the potential well
$2/\mu$ to be sufficiently small, or by fixing $2/\mu$ and choosing
$r_s$ to be sufficiently large.  We have tried both of these
approaches, but did not obtain a smooth variation of the energy when
reducing the range of the interaction, at least partly because the
wave function optimization becomes more difficult.  We obtained
smoother results when increasing the value of $r_s$ while keeping $\mu
= 12$, as shown in Table \ref{Universal_param_with_density}, and the
optimizations worked well in these calculations.  The value of $\xi$
slowly increases with $\mu r_s$, but this behavior could arise from
the fixed-node error inherent in the trial wave functions.  Note that
the differences between the VMC and DMC energies increase with $\mu
r_s$, indicating that the trial wave functions are becoming less
accurate.  Increasing the size of the simulation cell extends the
range of the trial wave function and makes it more difficult to
represent.

\begin{table}[!ht]
{\centering \begin{tabular}{lcc}
\hline\hline
$\mu r_s$  & $\xi$ (VMC) & $\xi$ (DMC) \\
\hline
12 & 0.456(1) & 0.4370(4) \\
14 & 0.462(1) & 0.4379(4) \\
16 & 0.470(1) & 0.4392(6) \\
18 & 0.477(1) & 0.4442(5) \\
\hline\hline
\end{tabular}\par}
\caption[]{The energy parameter, $\xi$, with $\mu = 12$ and different
particle density parameters $r_s$.}
\label{Universal_param_with_density}
\end{table}

\section{Conclusions}
\label{sec:Conclusions}

We have performed VMC and DMC calculations of the atomic Fermi gas at
zero temperature with a short ranged interaction at unitarity using
128 particles, which is larger than in previous calculations.  
Our DMC result of $\xi = 0.4339(1)$ is similar to previous DMC
results,\cite{Astrakharchik:PRL:2004,Chang:PRL:2005,Carlson:PRL:2005}
but is significantly larger than that of $\xi = 0.360(20)$ obtained
from a recent application of the epsilon
expansion.\cite{Nishida:PRA:2009} The VMC and DMC results for the
spherically-averaged pair-correlation functions, the momentum density,
and the one-body density matrix were in good agreement, illustrating
the high accuracy of our trial wave functions.  The VMC and DMC
results for the two-body density matrix, and the condensate fraction
derived from it, are somewhat different, indicating that significant
errors still arise from the approximate trial wave functions and/or
finite size effects.  We have calculated a somewhat smaller condensate
fraction than in other studies using similar methods.  We also
calculated the variation of $\xi$ with particle density for a fixed
well width, finding a relatively small variation over the range of
densities studied, but we were unable to draw a firm conclusion as to
whether these represent real variations with well width or whether
they are due to variations in the fixed-node errors.

\begin{acknowledgments}
We are grateful to Neil Drummond for fruitful discussions.  
This work was supported by the Engineering and Physical Sciences
Research Council (EPSRC) of the UK\@.  Computational resources were
provided by the Cambridge High Performance Computing Service.
\end{acknowledgments} 

\bibliographystyle{h-physrev}

\begin{thebibliography}{10}
\bibitem{units} We use Hartree atomic units, so that energies are in
Ha and masses are in multiples of the electron mass.
\end{thebibliography}

\end{document}